\documentclass[prd,twocolumn,nofootinbib,superscriptaddress]{revtex4-2}

\usepackage{preamble}

\begin{document}

\title{The optical force on a dielectric slab
}

\author{Andrew Laeuger\,\orcidlink{0000-0002-8212-6496}}
\email{alaeuger@caltech.edu}
\affiliation{Department of Physics and Astronomy, Caltech, Pasadena, California 91125, USA}

\author{Kevin Kuns\,\orcidlink{0000-0003-0630-3902}}
\affiliation{LIGO Laboratory, Department of Physics, Massachusetts Institute of Technology, Cambridge, MA 02139, USA}

\author{Yanbei Chen\,\orcidlink{0000-0002-9730-9463}}
\affiliation{Department of Physics and Astronomy, Caltech, Pasadena, California 91125, USA}

\author{Nancy Aggarwal\,\orcidlink{0000-0002-0026-3877}}
\affiliation{Department of Physics and Astronomy, University of California, Davis, California 95616, USA}

\date{\today}

\begin{abstract}
    Radiation-pressure forces on transmissive optics, including dynamical backaction, are usually computed from the momentum flux of the external fields. For a dielectric of finite thickness, this omits the time dependence of the electromagnetic momentum stored inside. We compute the force on a lossless dielectric slab of arbitrary thickness $L_s$ submerged in a dielectric fluid, proceeding from both Lorentz force on its bound charges and from the external momentum fluxes for arbitrary carrier and sideband fields. We treat the surrounding medium through a prescribed linear dielectric response and do not model the accompanying mechanical response of the fluid, including pressure and electrostriction. These two methods both recover the expected force in the limit of a thin reflector, but differ in the more general case by the rate of change of the field momentum stored in the slab. This storage term vanishes for static fields and at low frequencies reduces to $(L_s/c^2)\,dP/dt$, where $P$ is the net power through the slab. In a cavity, where $P$ depends on the slab position, the storage term produces a velocity-dependent force. The Abraham and Minkowski momentum formulations applied to the storage term are reconciled to give an unambiguous force on the dielectric in the case of an exterior vacuum environment, where there is no fluid response to model.

\end{abstract}

\maketitle

\ssec{Introduction}
Accurately describing the optical force on dielectric elements is essential for modeling optomechanical systems, from tethered membranes \cite{Thompson2008, Jayich2008, Nair2016} and levitated plates \cite{Arvanitaki2013, Aggarwal2022} to the thick transmissive optics of gravitational-wave detectors \cite{LIGOScientific:2014pky}. These forces, including dynamical backaction \cite{Aspelmeyer2014}, are usually computed from the momentum flux of the fields outside the element \cite{CorbittFramework}, an approach that underlies experiments using radiation pressure to control optomechanical cavities, measure quantum back action, and generate squeezed light \cite{CripeLocking, CripeQRPN, AggarwalSqueezing}. For subwavelength particles, the point-dipole picture of gradient and scattering forces applies \cite{Gordon, GordonplusAshkin, Ashkin:1978zsu, Ashkin:86, ChangSpheres}, while for wavelength-scale particles in the Mie regime the force follows from the full three-dimensional scattered field, typically for monochromatic illumination. The planar elements considered here differ from both: their lateral extent exceeds the beam waist, so they scatter only into the forward and backward modes and become part of the cavity rather than a small perturbation to it. The problem then reduces to one dimension, where the dependence on thickness relative to the optical wavelength is captured by standard transfer-matrix methods for arbitrary thickness. Because these elements sit inside cavities whose fields respond to their motion, the fields are inherently time dependent, and the propagation delay across the element, compared with the timescale of the field fluctuations, becomes an additional scale.

In this work, we derive the optical force on a planar dielectric slab from an incident plane electromagnetic wave, retaining the full time dependence of the electromagnetic field inside the dielectric. Our treatment allows for arbitrary slab thickness and time-dependent cavity fields and is formulated in terms of the carrier and sideband fields relevant to linearized cavity optomechanics. The planar geometry permits the internal fields and the resulting force to be obtained analytically. Furthermore, it is motivated by the need for self-consistent optomechanical analyses of tethered membranes and levitated dielectric slabs inserted into Fabry-Perot cavities and whose cross sections capture the full optical mode profile \cite{DielectricLetter,FormalismPaper}, an area of growing experimental interest \cite{Winstone2022}. Similar calculations have been carried out in past works, obtaining identical results to ours in the proper limits, though they have generally focused on the physical consequences in the case of distinct electromagnetic pulses rather than continuous waves \cite{Loudon:06, Cheung:2011izy, Balazs1953, Mansuripur:04}.

The force can be calculated either from the Lorentz force acting on the dielectric or from the net electromagnetic momentum flux through a surface surrounding it. The force-density and external momentum-flux methods generate different results, but the ambiguity is reconciled precisely after accounting for the field momentum stored within the dielectric. The storage term vanishes at DC and is suppressed by $\Omega L_s/c$ relative to the conservative (radiation-pressure) force. In a cavity it nevertheless contributes a velocity-dependent force at lowest order in $\Omega$, corresponding to an additional mechanical damping rate $\Delta\Gamma = (L_s/mc^2)\,\partial P_\text{net}/\partial x$, where $P_\text{net}$ is the net power flowing through the slab at position $x$. A velocity-dependent force has also been identified for this system in \cite{Castanos2015}, though its origin is identified as velocity-driven polarization of the slab rather than time-dependence of the vacuum electromagnetic momentum stored within the slab. For typical experimental setups, the mechanical damping identified here is likely weak relative to the cavity-induced damping, though it may become more relevant for high finesse cavities, where the cavity-induced damping decreases, or for setups where the phase accumulation within the slab is a substantial fraction of that over the entire cavity, i.e., $nL_s\gtrsim L_c$.

This correction term also provides a direct relation to the longstanding Abraham--Minkowski controversy concerning electromagnetic momentum in dielectric media. These two formulations assign different momenta to the electromagnetic field in matter, but modern treatments have restored the agreement between observable forces and momentum conservation by recognizing that each formulation corresponds to a different partition of the total momentum between electromagnetic and material degrees of freedom \cite{Brevik1979,Pfeifer2007,Barnett2010,BarnettLoudon2010}. The internal momentum-storage term provides a particularly direct realization of this principle for a finite dielectric slab: the force obtained by momentum flux balance at each interface of the slab becomes consistent under the two formulations after recognizing that the material contribution cannot cross the slab boundary. The same bookkeeping extends to a slab immersed in a dielectric environment of index $n_2$: the force acquires a correction proportional to $\chi_{e,2}$, which in the high-reflectivity limit recovers the known $n_2$-fold enhancement of radiation pressure on an immersed reflector \cite{Jones1954, Jones1978TheMO}, but which does not reduce to either formulation for a general slab.

\ssec{Preliminaries}
\label{sec: quadrature formalism preliminaries}
Throughout this work we consider a linear, nondispersive, nonmagnetic, lossless, and uncharged dielectric with spatially uniform refractive index $n_1$, immersed in a dielectric environment of index $n_2$. We assume an electric field $\vec{\mathbb{E}}$ which is polarized purely in the $z-$direction, tangential to the dielectric surfaces:
\begin{equation}
    \vec{\mathbb{E}}(x,y,z,t)=f(y,z)\mathcal{E}(x,t)\hat z,
\end{equation}
where $f(y,z)$ is the cross-sectional distribution of the field amplitude. We take the reflecting surfaces to have radii larger than the beam waist and the beam profile to not vary significantly over the length of the slab. Thus, $f(y,z)$ only produces a prefactor $A_\text{eff}=\int |f(y,z)|^2\,dy\,dz$ in the force, an effective area found by integrating the intensity profile over the slab cross section.

We elect to employ the two-photon formalism, which was initially outlined in \cite{CavesAndSchumaker1,CavesAndSchumaker2} and further developed in \cite{CorbittFramework}; we have extended it to incorporate planar dielectrics in \cite{DielectricLetter, FormalismPaper}. Within this formalism, we first split $\mathcal{E}$ into a component propagating in the $+x$ direction ($\mathcal{E}_+$) and a component propagating in the $-x$ direction ($\mathcal{E}_-$). We further decompose $\mathcal{E}_{\pm}(x,t)$ as 
\begin{equation}
    \mathcal{E}_\pm(x,t)=\mathcal{E}_{\pm,1}(x,t)\cos(\omega_0 t)+\mathcal{E}_{\pm,2}(x,t)\sin(\omega_0 t),
\end{equation}
where $\omega_0$ is the carrier frequency of the field and $\mathcal{E}_{\pm,1/2}$ are the amplitude and phase quadrature components of the field. It is useful to define
\begin{equation}
    \vec{\mathcal{E}}_\pm=\begin{pmatrix}
        \mathcal{E}_{\pm,1} \\ \mathcal{E}_{\pm,2}
    \end{pmatrix},
\end{equation}
which will reduce the complexity of future expressions.

The two-photon formalism is particularly useful here because it separates the DC and AC components of the fields. The two quadratures are broken into a constant and fluctuating component, i.e.,
\begin{equation}
    \mathcal{E}_{\pm,1/2}(t)=E_{\pm,1/2}+e_{\pm,1/2}(t).
\end{equation}
We then write $e_{\pm,1/2}(\Omega)$ as the Fourier transform of $e_{\pm,1/2}(t)$, using the convention
\begin{equation}
    e_{\pm,1/2}(\Omega)\equiv \int_{-\infty}^\infty e_{\pm,1/2}(t)\exp(i\Omega t)dt.
\end{equation}
The constant part gives the carrier (DC) field amplitude, while the component oscillating at $\Omega$ is the sideband (AC) field.  The time-evolution of the quadrature components is assumed to be slow relative to $\omega_0$. Much like the quadrature vector $\vec{\mathcal{E}}_\pm$ defined earlier, we write
\begin{equation}
    \vec E_\pm=\begin{pmatrix}
        E_{\pm,1} \\ E_{\pm,2}
    \end{pmatrix},\quad 
    \vec e_\pm(\Omega)=\begin{pmatrix}
        e_{\pm,1}(\Omega) \\ e_{\pm,2}(\Omega)
    \end{pmatrix}.
\end{equation}

\begin{figure}
    \centering
    \includegraphics[width=0.7\linewidth]{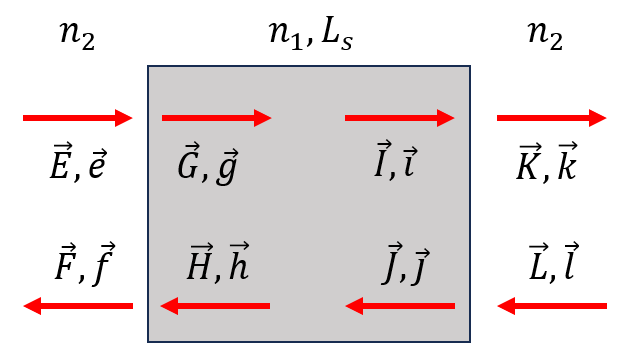}
    \caption{Schematic diagram of the carrier and sideband fields on either side of the two interfaces which bound the slab of thickness $L_s$ and index of refraction $n_1$. The exterior medium has index of refraction $n_2$.}
    \label{fig: dielectric slab}
\end{figure}

Our model appears in
Fig. \ref{fig: dielectric slab}. Instead of describing the right- and left-propagating fields at each location with $x$ and a $\pm$ subscript, we do so with distinct letter quadrature vectors. The paired letter vectors, which match the notation of \cite{FormalismPaper}, describe the fields on either side of each slab-environment interface, where $\vec E,\vec G,\vec I,\vec K$ (and their sideband counterparts) propagate in the $+x$ direction and $\vec F,\vec H,\vec J,\vec L$ in the $-x$ direction.

The exact propagation relations between the different fields are not needed to carry out the forthcoming calculation. However, the propagation relations inside the slab are useful for understanding some of the context in which the results are placed. These relations are derived in detail in \cite{FormalismPaper}; in short, for the carrier and sidebands:
\begin{subequations}
\label{eq: moving dielectric fields}
\begin{alignat}{2}
    \vec H &= \mathbb{R}(n_1kL_s)\vec J, \\
    \vec I &= \mathbb{R}(n_1kL_s)\vec G, \\
    \vec h &= \mathbb{R}(n_1kL_s)\exp(in_1\Omega L_s/c)\vec j, \\
    \vec i &= \mathbb{R}(n_1kL_s)\exp(in_1\Omega L_s/c)\vec g,
\end{alignat}
\end{subequations}
where $k=\omega_0/c$ is the carrier wavevector, $L_s$ is the slab thickness, and $\mathbb{R}$ is the $2\times2$ rotation matrix.

\ssec{Momentum balance and force density}
A general local flux balance takes the form
\begin{equation}
\frac{\partial \vec{\mathbbm{g}}}{\partial t}+\vec{\mathbbm{f}}=\nabla\cdot\sigma,
\label{eq: general flux balance}
\end{equation}
where $\vec{\mathbbm{g}}$ and $\vec{\mathbbm{f}}$ are the momentum and force densities and $\sigma$ is the stress tensor. For electromagnetic fields in vacuum, the momentum density is $\vec{\mathbbm{g}}_{\rm vac}=\epsilon_0\vec{\mathbb{E}}\times\vec{\mathbb{B}}$ and the stress tensor is
\begin{multline}
\sigma_{\text{vac},ij}\equiv \sigma_{\text{vac},ij}^{E}+\sigma_{\text{vac},ij}^{B}
\\=\epsilon_0\left(\mathbb{E}_i\mathbb{E}_j-\frac12\delta_{ij}\mathbb{E}^2\right)
+\frac{1}{\mu_0}\left(\mathbb{B}_i\mathbb{B}_j-\frac12\delta_{ij}\mathbb{B}^2\right).
\end{multline}
The corresponding force density is the total of that on the free and bound charges -- though with no free charges, $\vec{\mathbbm{f}}_\text{free}=0$. Although $\vec{\mathbbm{g}}_\text{vac}$ and $\sigma_\text{vac}$ take their vacuum forms, this balance still holds inside the dielectric. At the microscopic level there is no medium, only charges moving in vacuum; averaging over a volume larger than the molecular spacing but smaller than $c/\omega_0$ replaces the microscopic fields by $\vec{\mathbb{E}}$ and $\vec{\mathbb{B}}$ and the molecular charge and current densities by the bound versions $\rho_b$ and $\vec{\mathbb{J}}_p$. The near-field forces between neighboring molecules omitted by this averaging cancel in pairs, so the integrated force density gives the net force on the rigid slab.

The polarization field in the linear dielectric slab is
\begin{equation}
\vec{\mathbb{P}}=\epsilon_0\chi_{e,1}\vec{\mathbb{E}},
\qquad
\chi_{e,1}=n_1^2-1,
\end{equation}
giving bound charge and polarization current densities of
\begin{equation}
\rho_{b}=-\nabla\cdot\vec{\mathbb{P}},
\qquad
\vec{\mathbb{J}}_{p}=\frac{\partial\vec{\mathbb{P}}}{\partial t}.
\end{equation}
Their Lorentz force density,
$\vec{\mathbbm{f}}_b=\rho_b\vec{\mathbb{E}} +\vec{\mathbb{J}}_p\times\vec{\mathbb{B}},$
can be rearranged using Maxwell's equations to separate the purely electromagnetic momentum from that associated with the polarization response.
\begin{equation}
\vec{\mathbbm{f}}_b
=
\chi_{e,1}\frac{\partial\vec{\mathbbm{g}}_\text{vac}}{\partial t}
-\chi_{e,1}\nabla\cdot\sigma^E_\text{vac}-\epsilon_0(\vec{\mathbb{E}}\cdot\nabla\chi_e)\vec{\mathbb{E}}.
\label{eq: bound force rearranged}
\end{equation}
The generic susceptibility $\chi_e$ should be understood as position-dependent, with said dependence suppressed. At the slab surfaces, $\chi_e$ is discontinuous, so the final term is a surface contribution. However, because the electric field is polarized parallel to the surfaces, $\vec{\mathbb{E}}\cdot\nabla\chi_e=0$, and thus the surface term vanishes. We substitute the vacuum flux balance law $\partial_t\vec{\mathbbm{g}}_\text{vac}=-\vec{\mathbbm{f}}_b+\nabla\cdot\sigma_\text{vac}$ into the bound current\footnote{The bound charge density $\rho_b=0$ in the bulk since the susceptibility is uniform and there are no free charges.} force density in Eq. \eqref{eq: bound force rearranged}, obtaining
\begin{equation}
    \vec{\mathbbm{f}}_b=\frac{\chi_{e,1}}{1+\chi_{e,1}}\nabla\cdot\sigma^B_\text{vac}.
    \label{eq: bound current force}
\end{equation}

\ssec{Computing the total mechanical force}
The total force on the bound charges and currents is obtained by integrating $\vec{\mathbbm{f}}_b$ over the dielectric volume. The construction in Eq. \eqref{eq: bound current force} was chosen to make the integration simple to perform via Gauss's Theorem, giving
\begin{equation}
    \mathbb{F}^\text{Lor}_x=\frac{\chi_{e,1}}{1+\chi_{e,1}}\frac{1}{2\mu_0}A_\text{eff}(\mathbb{B}_\text{left}^2-\mathbb{B}_\text{right}^2),
    \label{eq: force from B}
\end{equation}
where $\vec{\mathbb{B}}_\text{left/right}$ are the magnetic fields at the left and right surfaces of the slab \cite{Cheung:2011izy}.

The tangential components of the electric and magnetic fields are continuous across the boundaries; since we assume the fields are perpendicular to the cavity axis everywhere, $\vec{\mathbb{B}}_\text{left/right}$ can be evaluated just inside or just outside the dielectric. The magnetic field differs from the electric field in magnitude by a factor of $\sqrt{(1+\chi_{e,1/2})\epsilon_0\mu_0}=n_{1/2}/c$, and the left and right propagating waves must have opposite signs of $\vec{\mathbb{B}}$ relative to $\vec{\mathbb{E}}$. The magnetic fields are thus given by

\begin{align}
    \vec{\mathbb{B}}_\text{left}(t)={}&-\frac{n_1}{c}\,(\vec G+\vec g(t)-\vec H-\vec h(t))\cdot\begin{pmatrix}\cos(\omega_0t)\\ \sin(\omega_0t)\end{pmatrix}\hat y, \\
    \vec{\mathbb{B}}_\text{right}(t)={}&-\frac{n_1}{c}\,(\vec I+\vec i(t)-\vec J-\vec j(t))\cdot\begin{pmatrix}\cos(\omega_0t)\\ \sin(\omega_0t)\end{pmatrix}\hat y.
\end{align}

After time averaging over the laser period, the DC force along the cavity axis is
\begin{equation}
    \mathbb{F}^\text{Lor}_{x,\text{DC}}=\frac{\epsilon_0A_\text{eff}\chi_{e,1}}{4}\Bigr((\vec G-\vec H)^2-(\vec I-\vec J)^2\Bigr),
    \label{eq: polarization force DC}
\end{equation}
while the AC force, with $\Omega$-dependence suppressed, is
\begin{multline}
    \mathbb{F}^\text{Lor}_{x,\text{AC}}(\Omega)=\frac{\epsilon_0A_\text{eff}\chi_{e,1}}{2}
    \\\times \Bigr((\vec G-\vec H)\cdot(\vec g-\vec h)-(\vec I-\vec J)\cdot(\vec i-\vec j)\Bigr)
    \label{eq: polarization force AC}.
\end{multline}
The tangential field continuity conditions of Maxwell's equations provide the first instance where the external dielectric environment $n_2$ enters the calculation:
\begin{align}
    \vec{I}+\vec{J} &= \vec{K}+\vec{L} \,,\quad n_1(\vec{I}-\vec{J}) = n_2(\vec{K} - \vec{L})\\
    \vec{G}+\vec{H} &= \vec{E}+\vec{F} \,,\quad n_1(\vec{G}-\vec{H}) = n_2(\vec{E} - \vec{F}),
\end{align}
plus equivalent forms after exchanging the carrier vectors for the sidebands. Inserting these continuity relations gives a form more amenable to physical interpretation:
\begin{align}
    \mathbb{F}^\text{Lor}_{x,\text{DC}}={}&\frac{\epsilon_0A_\text{eff}}{2}\Bigl(n_2^2(\vec E^2+\vec F^2-\vec K^2-\vec L^2)\nonumber\\
    &\quad\quad\quad-(\vec G^2+\vec H^2-\vec I^{\,2}-\vec J^{\,2}) \nonumber\\
    &\quad-\frac{\chi_{e,2}}{2}\bigl((\vec E+\vec F)^2-(\vec K+\vec L)^2\bigr)\Bigr), \label{eq: DC pol force}\displaybreak[0]\\
    \mathbb{F}^\text{Lor}_{x,\text{AC}}={}&\epsilon_0A_\text{eff}\Bigl(n_2^2(\vec E\cdot\vec e+\vec F\cdot\vec f-\vec K\cdot\vec k-\vec L\cdot\vec l) \nonumber\\
    &\quad\quad\quad-(\vec G\cdot\vec g+\vec H\cdot\vec h-\vec I\cdot\vec i-\vec J\cdot\vec j)\nonumber\\
    &\hspace{-3em}-\frac{\chi_{e,2}}{2}\bigl((\vec E+\vec F)\cdot(\vec e+\vec f)-(\vec K+\vec L)\cdot(\vec k+\vec l)\bigr)\Bigr). \label{eq: AC pol force}
\end{align}

Alternatively, we can compute the momentum flux into a surface just enclosing the slab, using the vacuum-form Maxwell stress evaluated outside the dielectric:
\begin{multline}
    \mathbb{F}^\text{flux}_x=\oint_A\sigma_\text{vac}\cdot d\vec A
    \\
    =\frac{A_\text{eff}}{2}\Bigr(\epsilon_0(\mathbb{E}_\text{left}^2-\mathbb{E}_\text{right}^2)+\frac{1}{\mu_0}(\mathbb{B}_\text{left}^2-\mathbb{B}_\text{right}^2)\Bigr).
\end{multline}
After time-averaging, we obtain the DC and AC forces
\begin{align}
    \mathbb{F}_{x,\text{DC}}^\text{flux}={}&\frac{\epsilon_0A_\text{eff}}{4}\Bigl((\vec E+\vec F)^2+n_2^2(\vec E-\vec F)^2 \nonumber\\
    &\hspace{-1em}-(\vec K+\vec L)^2-n_2^2(\vec K-\vec L)^2\Bigr),\\
    \mathbb{F}_{x,\text{AC}}^\text{flux}={}&\frac{\epsilon_0A_\text{eff}}{2}\Bigl((\vec E+\vec F)\cdot(\vec e+\vec f)+n_2^2(\vec E-\vec F)\cdot(\vec e-\vec f) \nonumber\\
    &\hspace{-1em}-(\vec K+\vec L)\cdot(\vec k+\vec l)-n_2^2(\vec K-\vec L)\cdot(\vec k-\vec l)\Bigr).
\end{align}
Substituting in the field continuity relations leads to
\begin{align}
    \mathbb{F}^\text{flux}_{x,\text{DC}}&=\mathbb{F}^\text{Lor}_{x,\text{DC}}+\frac{\epsilon_0A_\text{eff}}{2}(\vec G^2+\vec H^2-\vec I^{\,2}-\vec J^{\,2}),\\
    \mathbb{F}^\text{flux}_{x,\text{AC}}&=\mathbb{F}^\text{Lor}_{x,\text{AC}}+\epsilon_0A_\text{eff}(\vec G\cdot\vec g+\vec H\cdot\vec h-\vec I\cdot\vec i-\vec J\cdot\vec j).
\end{align}
The difference between $\mathbb{F}^\text{flux}_x$ and $\mathbb{F}^\text{Lor}_x$ must be captured by the rate of change of the momentum stored within the slab, as required by momentum conservation. Applying the flux balance law, Eq. \eqref{eq: bound force rearranged}, and Gauss' Theorem, the stored vacuum field momentum $\vec{\mathbbm{p}}^\text{vac}$ obeys
\begin{equation}
\begin{aligned}
    \frac{d\vec{\mathbbm{p}}^\text{vac}}{dt}
    &= \int \frac{\partial \vec{\mathbbm{g}}_{\rm vac}}{\partial t} dV=\int_\text{slab} \nabla\cdot\Bigr(\sigma^E_\text{vac}+\frac{\sigma^B_\text{vac}}{1+\chi_{e,1}}\Bigr) dV
    \\
    &=\frac{A_\text{eff}}{2}\Bigr(\epsilon_0(\mathbb{E}_\text{left}^2-\mathbb{E}_\text{right}^2)+\frac{\mathbb{B}_\text{left}^2-\mathbb{B}_\text{right}^2}{\mu_0(1+\chi_{e,1})}\Bigr),
    \end{aligned}
\end{equation}
which after time averaging over $2\pi/\omega_0$ gives
\begin{align}
    \frac{d\mathbbm{p}^\text{vac}_{x,\text{DC}}}{dt}&=\frac{\epsilon_0A_\text{eff}}{2}(\vec G^2+\vec H^2-\vec I^{\,2}-\vec J^{\,2}),\\
    \frac{d\mathbbm{p}^\text{vac}_{x,\text{AC}}}{dt}&=\epsilon_0A_\text{eff}(\vec G\cdot\vec g+\vec H\cdot\vec h-\vec I\cdot\vec i-\vec J\cdot\vec j), \label{eq: time dependent EM storage}
\end{align}
making it readily apparent that
\begin{equation}
    \mathbb{F}^\text{flux}_{x,\text{DC/AC}}=\mathbb{F}^\text{Lor}_{x,\text{DC/AC}}+\frac{d\mathbbm{p}^\text{vac}_{x,\text{DC/AC}}}{dt}.
\end{equation}

At DC, $\vec I$ and $\vec H$ are just rotations of $\vec G$ and $\vec J$, respectively (cf. Eq. \eqref{eq: moving dielectric fields}), so $\mathbb{F}_{x,\text{DC}}^\text{flux}=\mathbb{F}_{x,\text{DC}}^\text{Lor}$ -- physically, all time-dependence is captured in the AC terms, so the time-dependence of the internal stored momentum cannot appear in the DC force. The difference at AC is nonzero because the sideband vectors accumulate a different phase compared to the carrier vectors as they propagate through the dielectric; it can be simplified to
\begin{equation}
        \frac{d\mathbbm{p}^\text{vac}_{x,\text{AC}}}{dt}=\epsilon_0A_\text{eff}(1-\exp(in_1\Omega L_s/c))(\vec G\cdot\vec g-\vec J\cdot\vec j).
\end{equation}
The directional power (say, from fields $\vec F$ and $\vec f$), time-averaged over an optical period, is equal to $P=n_2\epsilon_0cA_\text{eff}\vec F^2/2$ at DC and $\delta P(\Omega)=n_2\epsilon_0cA_\text{eff}(\vec F\cdot\vec f(\Omega))$ at AC. Therefore, for small $n_1\Omega L_s/c$, the rate of change of stored momentum is $d\mathbbm{p}^\text{vac}_x/dt=(L_s/c^2)\,dP_\text{net}/dt$,  where $P_\text{net}$ is the net power through the slab, whose (linearized) fluctuating part is $n_1\epsilon_0cA_\text{eff}(\vec G\cdot\vec g-\vec J\cdot\vec j)$. In a cavity, $P_\text{net}$ also depends on the slab position \cite{DielectricLetter}; for mechanical frequencies well below the cavity linewidth, $dP_\text{net}/dt$ then contains the term $(\partial P_\text{net}/\partial x)\dot x$, so relative to $\mathbb{F}^\text{flux}_x$, $\mathbb{F}^\text{Lor}_x$ acquires a velocity-dependent force $-(L_s/c^2)(\partial P_\text{net}/\partial x)\dot x$, corresponding to a damping rate $\Delta\Gamma = (L_s/mc^2)\,\partial P_\text{net}/\partial x$. Once $n_1\Omega L_s/c\sim 1$, the internal cavity dispersion noticeably modifies the AC force.

\ssec{Slab force from the continuum perspective}
We now address this problem from the perspective of the constitutive response of the dielectric materials, which naturally gives way to an analysis which can invoke either the Abraham or Minkowski formulations of the momentum flux balance.
Beginning from Eq. \eqref{eq: bound force rearranged} and defining
\begin{equation}
\vec{\mathbbm{g}}_{\rm pol}
\equiv\epsilon_0\chi_{e,1}\vec{\mathbb{E}}\times\vec{\mathbb{B}}
=\chi_{e,1}\vec{\mathbbm{g}}_{\rm vac},
\end{equation}
the familiar momentum density in matter is
\begin{equation}
\vec{\mathbbm{g}}^M_{\rm matter}
\equiv\vec{\mathbbm{g}}_{\rm vac}+\vec{\mathbbm{g}}_{\rm pol}
=\epsilon_0(1+\chi_{e,1})\vec{\mathbb{E}}\times\vec{\mathbb{B}}.
\label{eq: effective Poynting and stress energy}
\end{equation}
Applying this definition to Eq. \eqref{eq: bound force rearranged}, plus the identity
\begin{equation}
    \chi_e\nabla\cdot\sigma^E_\text{vac}=\nabla\cdot(\chi_e\sigma^E_\text{vac})-\epsilon_0(\vec{\mathbb{E}}\cdot\nabla\chi_e)\vec{\mathbb{E}}+\epsilon_0\vec{\mathbb{E}}^2\nabla\chi_e/2,
\end{equation}
we obtain the Minkowski formulation of the flux balance
\begin{equation}
\frac{\partial\vec{\mathbbm{g}}^M_{\rm matter}}{\partial t}
+\vec{\mathbbm{f}}^M
=\nabla\cdot\sigma_{\rm matter},
\label{eq: free charges force}
\end{equation} 
where the matter stress tensor is $\sigma_{\rm matter}=(1+\chi_{e,1})\sigma^E_\text{vac}+\sigma^B_\text{vac}$ and the force density is $\vec{\mathbbm{f}}^M=\vec{\mathbbm{f}}_\text{free}-\epsilon_0\vec{\mathbb{E}}^2\nabla\chi_e/2$. With no free charges in the system, integrating this force density over a volume which encloses the slab just returns two surface terms, which can be rearranged using the continuity relations:
\begin{align}
    \mathbb{F}_{x,\text{DC}}^M={}&\frac{\epsilon_0A_\text{eff}}{2}\Bigl(n_2^2(\vec E^2+\vec F^2-\vec K^2-\vec L^2) \nonumber\\
    &\quad-n_1^2(\vec G^2+\vec H^2-\vec I^{\,2}-\vec J^{\,2})\Bigr), \label{eq: DC Minkowski force}\displaybreak[0]\\
    \mathbb{F}_{x,\text{AC}}^M={}&\epsilon_0A_\text{eff}\Bigl(n_2^2(\vec E\cdot\vec e+\vec F\cdot\vec f-\vec K\cdot\vec k-\vec L\cdot\vec l) \nonumber\\
    &\quad-n_1^2(\vec G\cdot\vec g+\vec H\cdot\vec h-\vec I\cdot\vec i-\vec J\cdot\vec j)\Bigr). \label{AC Minkowski force}
\end{align}

We can also consider this balance from the Abraham perspective, where the stress tensor is the same but now $\vec{\mathbbm{g}}_\text{matter}^A=\vec{\mathbbm{g}}_\text{vac}=\epsilon_0\vec{\mathbb{E}}\times\vec{\mathbb{B}}$ and the force density picks up an additional contribution, $\vec{\mathbbm{f}}^A=-\epsilon_0\vec{\mathbb{E}}^2\nabla\chi_e/2+\epsilon_0\chi_{e,1}\partial_t(\vec{\mathbb{E}}\times\vec{\mathbb{B}})$. After time-averaging over the optical period $2\pi/\omega_0$, this additional contribution only enters at AC. Indeed, including the integral of the term proportional to $\partial_t(\vec{\mathbb{E}}\times\vec{\mathbb{B}})$ in $\vec{\mathbbm{f}}^A$, we find
\begin{align}
    \mathbb{F}_{x,\text{DC}}^A={}&\frac{\epsilon_0A_\text{eff}}{2}\Bigl(n_2^2(\vec E^2+\vec F^2-\vec K^2-\vec L^2) \nonumber\\
    &\quad-(\vec G^2+\vec H^2-\vec I^{\,2}-\vec J^{\,2})\Bigr), \label{eq: DC Abraham force}\\
    \mathbb{F}_{x,\text{AC}}^A={}&\epsilon_0A_\text{eff}\Bigl(n_2^2(\vec E\cdot\vec e+\vec F\cdot\vec f-\vec K\cdot\vec k-\vec L\cdot\vec l) \nonumber\\
    &\quad-(\vec G\cdot\vec g+\vec H\cdot\vec h-\vec I\cdot\vec i-\vec J\cdot\vec j)\Bigr). \label{AC Abraham force}
\end{align}
with the second line of the Minkowski and Abraham DC forces both vanishing as previously stated.

\ssec{Discussion}
Through the Lorentz, Minkowski, and Abraham formulations, we have identified three unique ``forces", with AC forms in Eqs. \eqref{eq: AC pol force}, \eqref{AC Minkowski force}, and \eqref{AC Abraham force}, respectively. We now discuss the physical origin of the differences between them and how they are reconciled into a single unambiguous force which is the relevant quantity for the slab's center-of-mass motion.

All three force equations begin with terms which clearly resemble radiation pressure forces balanced at the two slab-exterior interfaces, with the Minkowski momentum ($P/c\times n$ for a wave of power $P$ in a medium of index $n$) used for the external terms and either the Minkowski or Abraham ($P/nc$) momenta used for the internal terms, while the Lorentz formulation acquires an additional correction term $\sim\chi_{e,2}$ from the external environment, i.e.
\begin{equation}
    \mathbb{F}^A_{x,\text{DC}}=\mathbb{F}^M_{x,\text{DC}}=\mathbb{F}^\text{Lor}_{x,\text{DC}}+\frac{\chi_{e,2}\epsilon_0A_\text{eff}}{4}[(\vec E+\vec F)^2-(\vec K+\vec L)^2].
\end{equation}
Because a strong reflector necessitates $\vec{\mathbb{E}}=0$ at its surface (i.e., $\vec F=-\vec E$ and $\vec L=-\vec K$), this term vanishes in the high reflectivity limit, in agreement with the established fact that the radiation pressure force on a reflecting surface suspended in a dielectric medium scales with $n$ times the incident power, or $n^2$ times the incident squared electric field amplitude \cite{Jones1954, Jones1978TheMO, ashkin1973radiation, LOUDON01042002}.

However, this term is not
generically zero for all slab configurations. The correction scales with $\vec{\mathbb{P}}\cdot\vec{\mathbb{E}}$ in the environment -- that is, its polarization energy density. Physically, it accounts for the portion of the incident light momentum which is transferred to the particles of the external polarizable fluid, instead of going into the slab. Neither the Abraham nor Minkowski formulations return a term like this, as under these prescriptions, the fluid polarization component is treated as part of the full electromagnetic momentum in the medium, and thus the entire piece which is subtracted in the Lorentz formulation is assumed to be transferred mechanically to the slab. When the exterior environment is vacuum, $\chi_{e,2}=0$, and the external field terms of all three formulations agree; when it is a fluid, the addition of the fluid pressure force to $\mathbb{F}^\text{Lor}_x$ is expected to achieve agreement with $\mathbb{F}^\text{A/M}_x$ at DC, assuming hydrostatic equilibrium is maintained. At AC, however, the exact amount of transferred momentum hinges nontrivially upon a consistent treatment of the fluid's density and pressure responses, including effects like electrostriction \cite{Brevik1979, Brevik2018}, which are beyond the scope of the present calculation.

In addition to the external terms, the internal terms also show a subtle difference, with the Minkowski formulation picking up an additional prefactor of $n_1^2$. This apparent discrepancy is resolved by carefully considering the difference in momentum partitions captured by the two formulations. We have seen that the Minkowski momentum flux density was defined such that the bound charge force was repackaged into the forms of $\vec{\mathbbm{g}}^M_\text{matter}$ and $\sigma_\text{matter}$ -- in other words, the Minkowski momentum flux density includes not just the pure vacuum electromagnetic component, but also the component due to the polarization of the linear dielectric\footnote{The pure electromagnetic and polarization components are described as ``optical" and ``material" contributions in \cite{Barnett2010}.}. Mathematically, we have
\begin{equation}
\begin{aligned}
    &\mathbb{F}^A_{x,\text{AC}}=\mathbb{F}^M_{x,\text{AC}}+\frac{d\mathbbm{p}_{x,\text{AC}}^\text{pol}}{dt}=\mathbb{F}^M_{x,\text{AC}}+\chi_{e,1}\frac{d\mathbbm{p}_{x,\text{AC}}^\text{vac}}{dt}.
\end{aligned}
\end{equation}

Because the polarization degrees of freedom are part of the slab itself, that component represents momentum exchange internal to the mechanical system rather than momentum contributing to its center-of-mass motion. The polarization momentum cannot cross the interface into the exterior environment, so the time derivative of the stored momentum which is actually able to leave the slab and influence its net mechanical recoil has been overcounted. The internal polarization momentum contribution exceeds the vacuum electromagnetic contribution by a factor of $\chi_{e,1}=n_1^2-1$; removing this piece from the (subtracted) stored momentum term recovers the agreement with the Abraham and Lorentz formulations.

Ultimately, the integrated Lorentz force density provides a straightforward route for computing the acceleration of the slab's center of mass. That being said, the balance of ingoing and outgoing electromagnetic momentum fluxes often provides a more intuitive perspective on the system; here, we have emphasized the necessity of carefully considering whether certain partitions of the momenta are able to influence the recoil of the dielectric when determining whether the Abraham or Minkowski formulations provide a more straightforward description. We have also provided an example where pure external fluxes fail to return the correct force, and additional physics, such as the momentum stored within the material, must be taken into account.

\ssec{Conclusion}
We have derived the force on a finite-thickness dielectric slab directly from the Lorentz force on its bound charges and currents and related the result to electromagnetic momentum balance. The bound charge force approach returns a different answer than balancing the total momentum flux into the slab at AC, but the difference is exactly accounted for in the time-dependence of the momentum stored in the fields within the slab. This internal storage contribution identically vanishes at DC; at AC, it enters at leading order as a velocity-dependent force which is $\sim n\Omega L_s/c$ times the conservative force, generating a damping term which becomes relevant when other sources of damping are small. 

More broadly, the result illustrates the importance of specifying the complete mechanical system when assigning electromagnetic momentum in matter. The Abraham--Minkowski literature emphasizes that alternative momenta can correspond to different partitions between field and material momentum, while observable forces remain fixed when the accompanying material contribution is treated consistently \cite{Pfeifer2007,Barnett2010,BarnettLoudon2010}. In the present problem, the Abraham and Minkowski formulations are resolved by considering how each partitions the momentum into parts which can pass through the bounding surface and parts which are necessarily trapped in the dielectric slab. The core distinction for when each formulation provides a more straightforward computation of the force lies in the role of the bound charges and currents: when the bound charges and currents are part of the moving object itself, the Abraham prescription gives the correct force from a pure momentum flux balance, and when they are external to the moving object, the Minkowski prescription becomes the correct choice.
This momentum-balance picture therefore is useful whenever the finite propagation time across a dielectric element cannot be neglected, and will soon be implemented in the gravitational wave detector simulation software \emph{Finesse} \cite{brown_2025_12662017}.

Assuming the fluid is in hydrostatic equilibrium with the slab, the force on the slab at DC is unambiguous:
\begin{equation}
    \mathbb{F}_{x,\text{DC}}=\frac{\epsilon_0A_\text{eff}}{2}n_2^2(\vec E^2+\vec F^2-\vec K^2-\vec L^2).
    \label{eq: DC force final}
\end{equation}
At AC, the internal contributions agree in all three partitions, but the external contributions fundamentally differ through their assumptions regarding how momentum passes from the fluid to the slab -- prescribing a fluid response model should in principle restore the agreement between Eqs. \eqref{eq: AC pol force} and \eqref{AC Abraham force}, which already holds when the external environment is vacuum, though that step lies beyond the present scope.

\ssec{Acknowledgments}
A.L. is grateful for support from the Fannie and John Hertz Foundation through a Hertz Fellowship. N.A. is supported by the UC Noyce Initiative and the Department of Energy award number DE-SC0009999. A.L. and Y.C. are supported by the Brinson Foundation, the Simons Foundation (Award Number 568762), and by US NSF Grants PHY--2309211 and PHY--2309231. K.K.\ is supported by NSF PHY--2309200, PHY--2309064, and PHY--2309267. We also thank Vivishek Sudhir and Dalziel Wilson for fruitful discussions throughout this work.

\bibliography{main}

\end{document}